\documentclass[12pt]{iopart}

\usepackage{amssymb}

\begin{document}

\title{Gauge Freedom in complex holomorphic systems}

\author{Carlos A. Margalli and J. David Vergara}
\address{Instituto de Ciencias Nucleares,\\
\it Universidad Nacional Aut\'onoma de M\'exico,\\
\it A. Postal 70-543 , M\'exico D.F., M\'exico}
\ead{carlos.margalli@nucleares.unam.mx}
\ead{vergara@nucleares.unam.mx}

\vspace{10pt}
\begin{indented}
\item[]january, 23  2017
\end{indented}

\begin{abstract}
The aim of this paper is to introduce and analyze a new gauge symmetry
that appears in complex holomorphic systems.
This symmetry allow us to project the system, using different gauge
conditions, to several real systems which are connect by gauge
transformations in the complex space. We prove that the space of solutions
of one system is related to the other by the gauge transformation. The
gauge transformations are in some cases canonical transformations.
However, in other cases are more general transformations that change the
symplectic structure, but there is still a map between the systems. In
this way our construction extend the group of canonical transformations in
classical mechanics. Also, we show how to extend the analysis to the
quantum case using path integrals by means of the
Batalin-Fradkin-Vilkovisky theorem and within the canonical
formalism, where we show explicitly that solutions of the Schr\"odinger
equation are gauge related.
\end{abstract}

\vspace{2pc}
\submitto{\JPA}

\section{Introduction}
Symmetries in physical systems play an important role in solving and
understanding  a system.
A very useful tool to describe local symmetries is the Dirac-Bergmann
theory of constrained Hamiltonian systems \cite{Dir, Hen}.
In this theory the local generators of gauge symmetries are first class
constraints, the existence of these constraints imply that the
dynamics of some degrees of freedom are not determined by the equations of
motion. However,
it is possible to determine the complete dynamics of the system by
removing the
remaining degrees of freedom by introducing a canonical gauge condition
that satisfy
certain requirements in combination with the first class constraints
\cite{Gribov}.
In addition, it is well known that the theories obtained by using
different gauge conditions are equivalent and are related by a gauge
transformation and the physics is gauge invariant.

On the other hand, in Quantum Mechanics and Quantum Field Theory is quite
natural to extend the symmetry transformations from the
real plane to the complex plane, and in fact, there has been attempts to
raise theories with physical sense
based on the structure of complex functions \cite{Ashtekar.b, Bender.1,
Moise}.

 In this work, we consider the extension to the complex plane of
any Lagrangian, with the condition that the extension is given only in
terms of holomorphic variables. This condition implies, that there exists a
hidden gauge symmetry in the
theory that allow us to project to the real physical space
\cite{Galileo, margalli:gnus}, in quite different ways, according to the
gauge choice. Furthermore,
these real theories are related under a gauge transformation in the
complex space.

Moreover, the extension on the complex plane of operators resulting from the
Quantum Mechanics is quite important considering the Hermiticity hypothesis.
Using the creation and annihilation operators, we can implement linear
canonical transformations  \cite{Moshinsky}
that map between unitary equivalent theories.  In our case, we extend this
concept since the gauge transformation that relates two
systems it is not necessarily canonical or unitary.  For example,  the
complex canonical transformation introduced by Ashtekar is in some sense
a very particular case of our work. Since, in  \cite{Morales} was shown
that the reality conditions can be interpreted has second class
constraints,
then this set of constraints can be interpreted as a pair of a first class
constraint that generates the gauge transformations and the corresponding
gauge fixing condition.

 In this paper, we analyze this gauge symmetry, we show that the temporal
evolution of the primary constraints is a manifestation of the
Cauchy-Riemann equations \cite{complex}, then if we
 impose that our Lagrangian is an holomorphic function, these equations
are satisfied identically and the primary constraints result a set of
generators of gauge transformations. As a second step
 we impose several sets of gauge conditions that imply different
projections to the real space. The resulting theories are not related by
a canonical transformation, however are related
 by a gauge transformation in the complex space and we show how from the
solution of one theory we can obtain the solution of the other gauge
related theory by using the complex gauge transformation. This gauge
transformation is implemented in the classical and the quantum cases.

The organization of this article is as follows, in Section  \ref{genertd},
 we consider a complex two-dimensional model,  and we show that there is a
trivial gauge
symmetry that leaves the system invariant. Using Dirac's method
\cite{Dir}, in the space separated in
real and imaginary parts we show that  associated with the symmetry there
are two first class constraints.  In consequence,
we can fix the gauge.  Using two different sets of gauge conditions we map
the initial complex space to two different real
systems. A case corresponds to two-dimensional harmonic oscillator and the
other to a Hamiltonian of a free particle in a non-constant metric
and a nontrivial symplectic structure. Solving the dynamics of both
systems we show that they are related by a complex gauge transformation in
the original complex phase space.

In Section \ref{or3} we introduce a gauge condition that lead us to a
non-commutative theory but this theory is gauge related
to the two-dimensional harmonic oscillator, with the usual symplectic
structure. We explicitly build the gauge transformations and we show how
the solutions of one system are connected with the solutions of the other
theory.  Section  \ref{brst} is devoted to analyze  the relation between
two gauge related quantum theories, to that end we use the BFV formalism
\cite{BFV} and we prove by using two different gauges  how we can go from
one theory to the other. In Section \ref{canonical} we implement our
procedure in the quantum canonical case, using a parametrized system. We
fix partially the gauge and  proof that the solutions of two gauge related
Schr\"odinger equations are gauge related. Finally Section \ref{conclu}
contains a brief discussion of our results.

\section{Gauge Conditions For The Case Of Two Dimensions
}\label{genertd}
The aim of this section is to describe the general basis of our procedure.
Specifically,
we consider an harmonic oscillator with complex variables in two
dimensions that has a gauge freedom. Fixing the gauge it is possible to
obtain a real system. The real system depends on the gauge conditions
selected and for our study
it is important to consider, the usual gauge conditions that reduce to a real
harmonic oscillator in two dimensions, the gauge conditions where we have
a linear dependence on momenta, and the gauge conditions
where we have quadratic dependence on momenta generating Dirac's brackets
for position variables that are  no trivial.

The complex model written in terms of a Lagrangian is
\begin{eqnarray}\label{Lagrangianacom}
L(z_1,z_2)=\frac{1}{2}\dot{z_1}^{2}+\frac{1}{2}\dot{z_2}^{2}-\frac{\omega_{1}^{2}}{2}z_1^{2}
-\frac{\omega_{2}^{2}}{2}z_2^{2}
\end{eqnarray}
with $z_1=x+iy$ and $z_2=u+iv$.
Here it is necessary to mention that the above Lagrangian is
invariant, if we apply the transformation
\begin{eqnarray}\label{finitet}
 x'=x+\lambda(t),\qquad y'=y+i\lambda(t).\nonumber\\
u'=u+\bar{\lambda}(t),\qquad v'=v+i\bar{\lambda}(t)
\end{eqnarray}
where $\bar{\lambda}(t),\lambda(t)$ are arbitrary functions of the time,
then we have a local symmetry. Notice that we are considering an extension
of the usual allowed transformations, since the real functions $y$ and $v$
get an imaginary part.  However, we obtain
\begin{equation}
 z'_{1}=z_{1},\qquad z'_{2}=z_{2}, \qquad L'=L,
\end{equation}
and in this way we get a legitimate symmetry. Now, we proceed to the
canonical analysis of this symmetry. From the above Lagrangian, we obtain
the momenta
\begin{eqnarray}
p_{x}=\dot{x}+i\dot{y},\qquad
p_{y}=-\dot{y}+i\dot{x},\\
p_{u}=\dot{u}+i\dot{v},\qquad
p_{v}=-\dot{v}+i\dot{u},\nonumber
\end{eqnarray}
from the definition of these momenta we observe that there are two primary
constraints
\begin{eqnarray}\label{constraingamma}
 \Phi_{0}=p_{x}+ip_{y}\approx 0,\\
 \Phi_{1}=p_{u}+ip_{v}\approx 0.\nonumber
\end{eqnarray}
On another hand, if we take the Lagrangian (\ref{Lagrangianacom}),
we compute the canonical Hamiltonian, using the ordinary Lagrange
transform, that results
\begin{eqnarray}\label{hamred}
 H_{2C}&=&p_x \dot x +p_y \dot y+p_u \dot u +p_v \dot v -L \\
 &=&\frac{1}{2}p_{x}^{2}+\frac{\omega_{1}^{2}}{2}(x+iy)^{2}+\frac{1}{2}p_{u}^{2}+\frac{\omega_{2}^{2}}{2}(u+iv)^{2}.\label{hamred1}
\end{eqnarray}
The symplectic structure of our system is given by the fundamental Poisson
brackets
\begin{eqnarray}
 \{x,p_{x}\}=1, \qquad  \{y,p_{y}\}=1, \qquad \{u,p_{u}\}=1, \qquad
\{v,p_{v}\}=1.
\end{eqnarray}
Notice that the Hamiltonian (\ref{hamred1}) is written only in terms of
the momenta $p_x$ and $p_u$, this is essentially due to the constraints
(\ref{constraingamma}), since the momenta $p_y$ and $p_v$ can be eliminated. The
total Hamiltonian for the system is
\begin{equation}
H_{2T}= H_{2C}+\mu^a \Phi_a,
\end{equation}
with $a=1,2$ and $\mu^a$ are arbitrary Lagrange multipliers. Using this
Hamiltonian the temporal evolution for the primary constraints are given
by
\begin{eqnarray}\label{phi0}
\dot{\Phi}_{0}=\{\Phi_{0},H_{2T}\}= 0, \qquad
\dot{\Phi}_{1}=\{\Phi_{1},H_{2T}\}= 0,
\end{eqnarray}
saying what the primary constraints are first class constraints and then
generators of gauge transformations. The gauge transformations of our
variables are given by
\begin{eqnarray}
\delta x&=&\{x, \epsilon^a \Phi_a\}= \epsilon^1, \qquad \delta y=\{y,
\epsilon^a \Phi_a\}= i\epsilon^1, \\
\delta u&=&\{u, \epsilon^a \Phi_a\}= \epsilon^2,\qquad \delta v=\{x,
\epsilon^a \Phi_a\}= i\epsilon^2,
\end{eqnarray}
Here we notice that these transformations are the infinitesimal form of
the finite Lagrange transformations (\ref{finitet}).
Due the gauge freedom in the system, we need to fix gauge conditions, our
first choice is
\begin{eqnarray}\label{gamma1}
 \gamma^{1}_{0}(y)=y\approx 0,\qquad \gamma^{1}_{1}(v)=v\approx 0,
\end{eqnarray}
and the brackets between the constraints are
\begin{eqnarray}
 \{\Phi_{0},\gamma^{1}_{0}\}=-i, \qquad \{\Phi_{1},\gamma^{1}_{1}\}=-i.
\end{eqnarray}
Now it is possible to make a complete set of second class constraints
$\chi_{a}=(\phi_{0},\gamma^{1}_{0}, \phi_{1},\gamma^{1}_{1},)$ with the
matrix of the second class constraints given by
\begin{equation}\label{am1}
\mathcal{A}_{ab}=\{\chi_{a},\chi_{b}\}=\left(
\begin{array}{rccccl}
0 & -i & 0 & 0\\
i & 0 &  0  & 0\\
0 & 0 & 0 & -i\\
0 & 0 & i & 0
\end{array}
\right).
\end{equation}
that has a determinanrt different of zero
\begin{eqnarray}
 \det{A}_{ab}=1
\end{eqnarray}
On another hand, we have a reduced Hamiltonian as a consequence of
applying the set of constraints
\begin{eqnarray}\label{17}
 H_{HO2}=\frac{1}{2}p_{\bar{x}}^{2}+\frac{\omega_{1}^{2}}{2}\bar{x}^{2}+\frac{1}{2}p_{\bar{u}}^{2}
+\frac{\omega_{2}^{2}}{2}\bar{u}^{2}.
\end{eqnarray}
Here, it is necessary to establish a change of variables in order to
avoid confusions.
For these ones and only for these conditions, we define:
\begin{eqnarray}
 x=\bar{x},\qquad y=\bar{y},
\end{eqnarray}
and now, we can compute the Dirac brackets  from
Hamiltonian theory that are reduced to
\begin{eqnarray}\label{19}
 \{\bar{x},p_{\bar{x}}\}^{*}=1,\qquad \{\bar{u},p_{\bar{u}}\}^{*}=1,
\end{eqnarray}
with other combinations of brackets, being zero and resulting
an harmonic oscillator in two dimensions.
Using these brackets it is straightforward to obtain the equations of
motion in usual way

\begin{eqnarray}\label{ecmov1}
\dot{\bar{x}}=\{\bar{x},H_{HO2}\}^{*}=p_{\bar{x}}, \qquad
\dot{\bar{u}}=\{\bar{u},H_{HO2}\}^{*}=p_{\bar{u}},\nonumber\\
\dot{p}_{\bar{x}}=\{p_{\bar{x}},H_{HO2}\}^{*}=-\omega_{1}^{2}\bar{x},
\qquad \dot{p}_{\bar{u}}=\{p_{\bar{u}},H_{HO2}\}^{*}
=-\omega_{2}^{2}\bar{u},
\end{eqnarray}
with solution given by

\begin{eqnarray}\label{haros}
 \bar{x}=A_{0}\sin{(\omega_{1}t+\delta_{0})},\qquad
\bar{u}=A_{1}\sin{(\omega_{2}t+\delta_{1})},
\end{eqnarray}
where it must be mentioned that the free parameters are fixed
with initial conditions, being $A_{0},\delta_{0},A_{1},\delta_{1}$.

In another hand, these gauge conditions are not special and they
could be selected in other way to obtain a different system, but both
theories
are gauge related on some region of the phase space.
A particular election is to consider the gauge conditions that include
linear terms of the momenta

\begin{eqnarray}\label{functions}
 \gamma_{0*}=y-ix+ig_{2}(x)p_{x}\approx 0,\nonumber\\
\gamma_{1*}=v-iu+ih_{2}(u)p_{u}\approx 0,
\end{eqnarray}
Then our set of second class constraints results $\chi_A=(\Phi_{0},
\gamma_{0*}, \Phi_{1}, \gamma_{1*})$.
By defining
\begin{equation}
 B=-ip_{x}\partial_{x}g_{2}(x),\qquad
C=-ip_{u}\partial_{u}h_{2}(u),
\end{equation}
in order to calculate easily the constraint matrix of second class
$\mathcal{B}_{AB}=\{ \chi_A,
\chi_B\}$. Associated to this set we obtain
\begin{equation}\label{am2}
\mathcal{B}_{ab}=\left(
\begin{array}{rccccl}
0 & B & 0 & 0\\
-B & 0 &  0  & 0\\
0 & 0 & 0 & C\\
0 & 0 & -C & 0
\end{array}
\right).
\end{equation}
The determinant for (\ref{am2}) must be different of zero  in order to get
a good set of constraints. At this point, it is necessary to consider a
concrete example of
(\ref{functions}) and to choose specifically the functions

\begin{eqnarray}\label{gh}
 g_{2}(x)=x, \qquad h_{2}(u)=u,
\end{eqnarray}
with an associated determinant for (\ref{am2}) and (\ref{gh})
\begin{equation}\label{determin}
 \det
B_{ab}=B^{2}C^{2}=(p_{x}\partial_{x}g_{2})^{2}(p_{u}\partial_{u}h_{2})^{2}=p_{x}^{2}p_{u}^{2}.
\end{equation}
On another hand, we obtain the reduced Hamiltonian using
these gauge conditions
\begin{eqnarray}\label{Hamiltonianred}
H_{cons}=[\frac{1}{2}+\frac{\omega_{1}^{2}}{2}g^{2}_{2}(x)]p_{x}^{2}
+[\frac{1}{2}+\frac{\omega_{2}^{2}}{2}h^{2}_{2}(u)]p_{u}^{2}\nonumber\\
=[\frac{1}{2}+\frac{\omega_{1}^{2}}{2}x^{2}]p_{x}^{2}
+[\frac{1}{2}+\frac{\omega_{2}^{2}}{2}u^{2}]p_{u}^{2}
=g^{\mu\nu}\frac{p_{\mu}p_{\nu}}{2},
\end{eqnarray}
where, we see that it is equivalent to a free particle in the metric
\begin{equation}
{g_{\mu \nu }} = \left( {\begin{array}{*{20}{c}}
{\frac{2}{{1 + \omega _1^2{x^2}}}}&0\\
0&{\frac{2}{{1 + \omega _2^2{u^2}}}}
\end{array}} \right),
\end{equation}
and from this expression we obtain that the curvature
$R=0$. However, the  dynamics of the system is not only given by the
Hamiltonian, we need to take into account that
the symplectic structure is given in the reduced space by the Dirac
brackets

\begin{eqnarray}\label{poisson}
 \{x,p_{x}\}^{*}=\frac{1}{p_{x}\partial_{x}g_{2}(x)}
 =\frac{1}{p_{x}}, \qquad
\{u,p_{u}\}^{*}=\frac{1}{p_{u}\partial_{u}h_{2}(u)}=\frac{1}{p_{u}
},
\end{eqnarray}
and zero for other combinations. Through the above brackets
(\ref{poisson}) we determine the temporal evolution for
the phase space variables
\begin{eqnarray}\label{EMxu}
 \dot{x}=\{x,H_{cons}\}^{*}=\frac{(1+\omega_{1}^{2}g_{2}^{2}(x))}{\partial_{x}g_{2}(x)}
=(1+\omega_{1}^{2}x^{2}),\nonumber\\
\dot{p}_{x}=\{p_{x},H_{cons}\}^{*}=-\omega_{1}^{2}g_{2}(x)p_{x}
=-\omega_{1}^{2}xp_{x},\nonumber\\
 \dot{u}=\{u,H_{cons}\}^{*}=\frac{(1+\omega_{2}^{2}h_{2}^{2}(u))}{\partial_{u}h_{2}(u)}
=(1+\omega_{2}^{2}u^{2}), \nonumber\\
\dot{p}_{u}=\{p_{u},H_{cons}\}^{*}=-\omega_{2}^{2}h_{2}(u)p_{u}=-\omega_{2}^{2}up_{u},
\end{eqnarray}
and we obtain an expression associated to the momenta in this reduced theory
\begin{eqnarray}
 p_{x}=p_{x0}\exp{(-\omega_{1}^{2}\int_{t_{0}}^{t}\!d\tau g_{2}(x))}
=p_{x0}\exp{(-\omega_{1}^{2}\int_{t_{0}}^{t}\!d\tau x)},\nonumber\\
 p_{u}=p_{u0}\exp{(-\omega_{2}^{2}\int_{t_{0}}^{t}\!d\tau h_{2}(u))}
=p_{u0}\exp{(-\omega_{2}^{2}\int_{t_{0}}^{t}\!d\tau  u)}.
\end{eqnarray}
Now it is possible to establish a reduced theory, as long as
we select the functions $g_{2}(x),h_{2}(u)$. Using (\ref{gh}) it is
possible to
solve the equations of motion and  obtain values for $p_{x}$  and $p_{u}$
once we have obtained solutions for $x$ and $u$.

The solutions obtained from equations (\ref{EMxu}) are
\begin{eqnarray}\label{sol1}
 x=\frac{1}{\omega_{1}}\tan{[\omega_{1}(t-t_{0})+\tan^{-1}(\omega_{1}x_{0})]},
\nonumber\\
u=\frac{1}{\omega_{2}}\tan{[\omega_{2}(t-t_{0})+\tan^{-1}(\omega_{2}u_{0})]},
\end{eqnarray}
and  for the momenta we get
\begin{eqnarray}\label{pceros}
 p_{x}=p_{x_{0}}(1+\omega_{1}^{2}x_{0}^{2})^{\frac{1}{2}}\cos[\omega_{1}(t-t_{0})+\tan^{-1}(\omega_{1}x_{0})],\nonumber\\
p_{u}=p_{u_{0}}(1+\omega_{2}^{2}u_{0}^{2})^{\frac{1}{2}}\cos[\omega_{2}(t-t_{0})+\tan^{-1}(\omega_{2}x_{0})].
\end{eqnarray}
On another hand, it is important to point out that there are points where
the mapping
does not apply. These points correspond to determinant (\ref{determin})
equal to zero, that imply that
squared momenta vanish (\ref{pceros}). Furthermore, the momenta are
periodic functions.

\subsection{Solutions Using Gauge Transformation}\label{contrano}

In this subsection, we will show you how to solve the above system through
a gauge transformation.
Let us consider the gauge conditions
(\ref{gamma1}) and (\ref{functions}),  then the difference between these
transformations parametrize the gauge transformation between the systems (\ref{17}) and (\ref{Hamiltonianred}) 
\begin{eqnarray}
\delta\gamma_{0}= \gamma_{0*}-\gamma^{1}_{0}=-ix+ig_{2}(x)p_{x}\approx
0,\nonumber\\
\delta\gamma_{1}=\gamma_{1*}-\gamma^{1}_{1}=-iu+ih_{2}(u)p_{u}\approx 0.
\end{eqnarray}
By applying these gauge transformations to our
variables we get
\begin{eqnarray}
 \delta x=x-\bar{x}=i\delta\gamma_{0}\{x,\Phi_0\},\qquad \delta
u=u-\bar{u}=i\delta\gamma_{1}\{u,\Phi_1\},\nonumber\\
\delta p_{x}=0,\qquad \delta p_{u}=0,
\end{eqnarray}
where, we can infer using (\ref{gh}),
\begin{eqnarray}\label{norm}
 \bar{x}=g_{2}(x)p_{x}=xp_{x},\qquad \bar{u}=h_{2}(u)p_{u}=up_{u},\qquad
{p}_{\bar x}=p_{x},\qquad {p}_{\bar u}=p_{u}.
\end{eqnarray}
Where we are considering that the variables with bar correspond to the
case (\ref{gamma1}) and the variables without bar to the gauge
(\ref{functions}).
Now, starting from the equations  (\ref{ecmov1})
\begin{eqnarray}
\dot{\bar{x}}=\frac{d}{dt}(xp_{x})=p_{\bar{x}}=p_{x},\qquad
\dot{\bar{u}}=\frac{d}{dt}(up_{u})=p_{\bar{u}}=p_{u},
\nonumber\\
\dot{p}_{\bar{x}}=\dot{p}_{x}=-\omega_{1}^{2}\bar{x},\qquad
\dot{p}_{\bar{u}}=\dot{p}_{u}=-\omega_{2}^{2}\bar{u},
\end{eqnarray}
by rewriting these equations in terms of variables without bar we get
\begin{eqnarray}
\dot{p}_{x}&=&-\omega_{1}^{2}xp_{x},\qquad
\dot{p}_{u}=-\omega_{2}^{2}up_{u},\label{momentos}\\
 \dot{x}&=&\frac{(1+\omega_{1}^{2}g_{2}^{2}(x))}{\partial_{x}g_{2}(x)}=(1+\omega_{1}^{2}x^{2}),
\dot{u}=\frac{(1+\omega_{2}^{2}h_{2}^{2}(u))}{\partial_{u}h_{2}(x)}=(1+\omega_{2}^{2}u^{2}),\label{posi1}
\end{eqnarray}
where it is important to mention that using the equations
(\ref{momentos}) we obtain equations (\ref{posi1}) through
 the gauge transformations (\ref{norm}). Furthermore, the equations
coincide with the equations (\ref{EMxu}).

Additionally, the solutions of both systems have a relationship
associated to the gauge transformation (\ref{norm}).
By applying to the  solution of harmonic oscillator (\ref{haros}), the gauge
transformation (\ref{norm}) we obtain
\begin{eqnarray}
x=\frac{\bar{x}}{p_{x}}=\frac{\sin{(\omega_{1}(t-t_{0})+\delta_{0})}}{\omega_{1}\cos{(\omega_{1}(t-t_{0})+\delta_{0})}}=
\frac{\tan{(\omega_{1}(t-t_{0})+\delta_{0})}}{\omega_{1}},  \nonumber\\
u=\frac{\bar{u}}{p_{u}}=\frac{\sin{(\omega_{2}(t-t_{0})+\delta_{1})}}{\omega_{2}\cos{(\omega_{2}(t-t_{0})+\delta_{1})}}=
\frac{\tan{(\omega_{2}(t-t_{0})+\delta_{1})}}{\omega_{2}},
\end{eqnarray}
that is equivalent to (\ref{sol1}) except a phase difference,
that is fixed by means of selecting a initial condition that
determines the interval where  is valid the solution.

We must notice that, this gauge transformation has determinant equal to
zero for a time given by
 $\frac{n\pi}{2\omega_{1,2}}$ when $n$ is odd,  see equation
(\ref{determin}).
This is linked  to  zones where the mapping is valid.

 On another hand, because first class constraints only depend from
momenta. Momenta from harmonic oscillator  coincide with momenta
 from the  theory resulting of  gauge conditions that are linear on
momenta (\ref{functions}).
However,  the initial conditions are different and it is possible to use
equations (\ref{momentos}) in order to find momenta (\ref{pceros}). With
respect to the solutions of both systems and the relationship between
initial conditions, we find
\begin{eqnarray}
 \delta_{0}=\tan^{-1}(x_{0}),\qquad \delta_{1}=\tan^{-1}(u_{0}),\nonumber\\
A_{0}=\frac{1}{\omega_{1}}p_{x_{0}}(1+\omega_{1}^{2}x_{0}^{2})^{\frac{1}{2}},\qquad
A_{1}=\frac{1}{\omega_{2}}p_{u_{0}}(1+\omega_{2}^{2}u_{0}^{2})^{\frac{1}{2}},
\end{eqnarray}
that are invertible in an temporal interval given by the tangent
function. In the next section we generalize this idea to other gauge
conditions.

\section{Quadratic Gauge Condition In Momenta}\label{or3}

Let's consider another kind of gauge conditions depending on quadratic
form of the momenta
\begin{eqnarray}\label{nornc}
 \gamma_{0+}=y-ix+ig_{2}(x)p_{x}p_{u},\nonumber\\
\gamma_{1+}=v-iu+ih_{2}(u)p_{u}.
\end{eqnarray}

As an important element of this work and in order to make easy the
notation, we define the next quantities
\begin{eqnarray}
 A_{1}=-ip_{x}p_{u}\partial_{x}g_{2}(x),\qquad
B_{1}=-g_{2}(x)p_{x}+p_{x}p_{u}g_{2}(x)\partial_{u}h_{2}(u),\nonumber\\
E_{1}=-ip_{u}\partial_{u}h_{2}(u),
\end{eqnarray}
 in such a way that we establish a set of second class constraints
$\chi_{a}=(\Phi_{0},\gamma_{0+},\Phi_{1},\gamma_{1+})$
 that allows us to find a constraint matrix by means of the Poisson
brackets  $\mathcal{G}_{ab}=\{\chi_{a},\chi_{b}\}$ resulting
\begin{equation}\label{am}
\mathcal{G}_{ab}=\left(
\begin{array}{rccccl}
0 & A_{1} & 0 & 0\\
-A_{1} & 0 &  0  & B_{1}\\
0 & 0 & 0 & E_{1}\\
0 & -B_{1} & -E_{1} & 0
\end{array}
\right).
\end{equation}
Furthermore the associated determinant  is
\begin{eqnarray}
 \det{\mathcal{G}_{ab}}=A_{1}^{2}E_{1}^{2}=p_{x}^{2}p_{u}^{4}[\partial_{x}g_{2}(x)]^{2}[\partial_{u}h_{2}(u)]^{2},
\end{eqnarray}
and it must be different from zero.

Here, we choose a concrete case and assign a value for $g_{2}$ and $h_{2}$
\begin{eqnarray}\label{gh2}
 g_{2}(x)=\frac{\theta}{2}x, \qquad h_{2}(u)=u,
\end{eqnarray}
that allows to find a value for the determinant
\begin{eqnarray}\label{det22}
 \det{(G_{ab})}=\frac{\theta^{2}}{4}p_{x}^{2}p_{u}^{4},
\end{eqnarray}
which depends from powers  of momenta. Furthermore, the reduced
Hamiltonian is given by
\begin{eqnarray}
H_{1Cons}=[\frac{1}{2}+\frac{\omega_{1}^{2}}{2}g^{2}_{2}(x)p_{u}^{2}]p_{x}^{2}
+[\frac{1}{2}+\frac{\omega_{2}^{2}}{2}h^{2}_{2}(u)]p_{u}^{2}
\end{eqnarray}
and using (\ref{gh2}) the Hamiltonian is
\begin{eqnarray}\label{49}
H_{1Cons}=[\frac{1}{2}+\frac{\omega_{1}^{2}}{2}(\frac{\theta}{2}x)^{2}p_{u}^{2}]p_{x}^{2}
+[\frac{1}{2}+\frac{\omega_{2}^{2}}{2}u^{2}]p_{u}^{2}.
\end{eqnarray}

Now, the symplectic structure of the system is given by the Dirac brackets
\begin{eqnarray}
\{x,p_{x}\}^{*}=\frac{1}{p_{x}p_{u}\partial_{x}g_{2}(x)}=\frac{2}{\theta
p_{x}p_{u}}, \nonumber\\
 \{x,u\}^{*}=\frac{g_{2}(x)}{p^{2}_{u}\partial_{x}g_{2}(x)\partial_{u}h_{2}(u)}
=\frac{x}{p_{u}^{2}},
\nonumber\\
\{u,p_{u}\}^{*}=\frac{1}{p_{u}\partial_{u}h_{2}(u)}=\frac{1}{p_{u}},
\label{DBNC1}
\end{eqnarray}
while the other brackets vanish. We observe from (\ref{DBNC1}) that in
this case the symplectic structure gives rise to a non-commutative theory.
In addition, we compute the temporal evolution of $x$ and $u$ resulting
\begin{eqnarray}
\dot{x}=\{x,H_{1Cons}\}^{*}=\nonumber\\
\frac{(1+\omega_{1}^{2}g_{2}^{2}(x)p^{2}_{u})}{\partial_{x}g_{2}(x) p_{u}}+
\omega_{2}^{2}xh_{2}(u)\partial_{u}h_{2}(u)
=\frac{2(1+\omega_{1}^{2}(\theta^{2}/4)x^{2}p^{2}_{u})}{\theta p_{u}}+
\omega_{2}^{2}xu,\nonumber\\
\dot{u}=\{u,H_{1Cons}\}^{*}=(1+\omega_{2}^{2}h_{2}^{2}(u))=(1+\omega_{2}^{2}u^{2}),\label{movi1}
\end{eqnarray}
and we observe that using the first equation it is possible to determine
the momentum  $p_{u}$, but from the second equation
we obtain a relationship that does not determine  $p_{x}$.
However, we can compute  $u(t)$ through  $h_{2}(u)=u$.

From the temporal evolution for $p_{x}$ and $p_{u}$,
we obtain  all the Hamilton equations of motion
\begin{eqnarray}
 \dot{p}_{x}=-\omega_{1}^{2}g_{2}(x)p_{x}p_{u}=-\frac{\omega_{1}^{2}}{2}\theta
xp_{x}p_{u} ,\qquad
\dot{p}_{u}=-\omega_{2}^{2}h_{2}(u)p_{u}=-\omega_{2}^{2}up_{u}. \label{movi2}
\end{eqnarray}

\subsection{Gauge Transformation with Non Trivial  Dirac's Brackets for
Positions}

Now, we are looking for the solution of the equations of motion
(\ref{movi1}) and (\ref{movi2}) by using the gauge transformations and its
relation with the harmonic oscillator problem.
We know that the parameters of the gauge transformations between the
gauges (\ref{gamma1}) and the
gauges  (\ref{nornc}) are given to first order by the differences
\begin{eqnarray}
 \delta\gamma_{0+}=\gamma_{0+}-\gamma_{0}=-ix+ig_{2}(x)p_{x}p_{u}=-ix+i\frac{\theta
x}{2}p_{x}p_{u},\nonumber\\
 \delta\gamma_{1+}=\gamma_{1+}-\gamma_{1}=-iu+ih_{2}(u)p_{u}=-iu+iup_{u}.
\end{eqnarray}
In accordance with the above, we establish the gauge transformations that
take us from the theory with Hamiltonian (\ref{17}) and symplectic
structure (\ref{19}) to the theory
with Hamiltonian (\ref{49}) with symplectic structure (\ref{DBNC1}), that
are given by
\begin{eqnarray}
 \delta x=\bar x -x=
-i\delta\gamma_{0+}\{x,\Phi\}=-x+g_{2}(x)p_{x}p_{u},\nonumber\\
\delta u=\bar u -u=-i\delta\gamma_{1}\{u,\Phi\}=-u+ih_{2}(u)p_{u},\nonumber\\
\delta p_{x}=0,\qquad \delta p_{u}=0, \label{gt6}
\end{eqnarray}
where we can deduce
\begin{eqnarray}\label{norm2}
 \bar{x}=g_{2}(x)p_{x}p_{u},\qquad \bar{u}=h_{2}(u)p_{u},\qquad
{p}_{\bar x}=p_{x},\qquad {p}_{\bar u}=p_{u}.
\end{eqnarray}
Then, we want to show that starting from the equations of motion for the
harmonic oscillator  (\ref{ecmov1}) and using the gauge transformation
(\ref{gt6}) that imply (\ref{norm2}), we can obtain the
equations of motion ( \ref{movi1}) and ( \ref{movi2}), where we use that
$g_{2}(x)=\frac{\theta x}{2}$ and $h_{2}(u)=u$,
resulting
\begin{eqnarray}\label{noconmutativaem}
\dot{p}_{x}=-\omega_{1}^{2}\bar{x}=-\omega_{1}^{2}\frac{\theta
x}{2}p_{x}p_{u},\qquad
\dot{p}_{u}=-\omega_{2}^{2}\bar{u}=-\omega_{2}^{2}up_{u},\\
\dot{x}=\frac{2(1+\omega_{1}^{2}(\theta^{2}/4)x^{2}p^{2}_{u})}{\theta p_{u}}+
\omega_{2}^{2}xu,\qquad
\dot{u}=(1+\omega_{2}^{2}u^{2}),\label{posi2}\label{momentos2}
\end{eqnarray}
So, starting from equations of motion of two oscillators with usual
symplectic structure, we arrive to a system of equations with non
commutative
variables. Similarly  to the case in Section 2, both systems have a
relationship
through a gauge transformation, in this case given by (\ref{norm2}). Now,
if we want to solve the equations (\ref{noconmutativaem}) and
(\ref{momentos2}), we use the solution
of the  harmonic oscillator (\ref{haros}), and apply the gauge
transformation  to obtain
\begin{eqnarray}
x=\frac{2\bar{x}}{\theta
p_{x}p_{u}}=\frac{2\sin{(\omega_{1}(t-t_{0})+\delta_{0})}}{\omega_{2}\omega_{1}\theta
\cos{(\omega_{1}(t-t_{0})+\delta_{0})}\cos{(\omega_{2}(t-t_{0})+\delta_{1})}}\nonumber\\
=
\frac{2\tan{(\omega_{1}(t-t_{0})+\delta_{0})}}{\omega_{1}\omega_{2}\theta\cos{(\omega_{2}(t-t_{0})+\delta_{1})}},
\nonumber\\
u=\frac{\bar{u}}{p_{u}}=\frac{\sin{(\omega_{2}(t-t_{0})+\delta_{1})}}{\omega_{2}\cos{(\omega_{2}(t-t_{0})+\delta_{1})}}=
\frac{\tan{(\omega_{2}(t-t_{0})+\delta_{1})}}{\omega_{2}},
\end{eqnarray}
that except a constant and a difference of phase, it coincides with the
solution of the system given by the equations (\ref{movi1}) and
(\ref{movi2}).

On the other side, we can fix the solutions to the initial conditions of
the non-commutative problem resulting

\begin{eqnarray}\label{noconsol1}
x=\frac{2\tan{(\omega_{1}(t-t_{0})+
\tan^{-1}(\frac{\omega_{1}\omega_{2}\theta x_{0}}
{2\sqrt{1+\omega_{2}^{2}u_{0}^{2}}}))}}{\omega_{1}\omega_{2}\theta\cos{(\omega_{2}(t-t_{0})+
\tan^{-1}(\omega_{2}u_{0}))}},
\nonumber\\
u=\frac{\tan{(\omega_{2}(t-t_{0})+\tan^{-1}(\omega_{2}u_{0}))}}{\omega_{2}},
\end{eqnarray}
For the momenta the equivalence of the initial conditions is given by

\begin{eqnarray}\label{noconsol2}
p_{x}=p_{x_{0}}(1+\frac{\omega_{1}^{2}\omega_{2}^{2}\theta^{2}x_{0}^{2}}{4(1+\omega_{2}^{2}u_{0}^{2})})^{\frac{1}{2}}
\cos(\omega_{1}(t-t_{0})+\tan^{-1}(\frac{\omega_{1}\omega_{2}\theta x_{0}}
{2\sqrt{1+\omega_{2}^{2}u_{0}^{2}}})),\nonumber\\
 p_{u}=
p_{u_{0}}(1+\omega_{2}^{2}u_{0}^{2})^{\frac{1}{2}}\cos(\omega_{2}(t-t_{0})
+\tan^{-1}(\omega_{2}u_{0})).
\end{eqnarray}
Furthermore, the relationship between initial conditions for both systems,
the harmonic oscillator in two dimensions (\ref{haros})
and the model with gauge condition non trivial, where Dirac's
brackets for positions are not commutative,
(\ref{noconsol1}) and (\ref{noconsol2}) is
\begin{eqnarray}
 \delta_{0}=\tan^{-1}(\frac{\omega_{1}\omega_{2}\theta
x_{0}}{2\sqrt{1+\omega_{2}^{2}}u_{0}^{2}}),\qquad
\delta_{1}=\tan^{-1}(\omega_{2}u_{0}),\nonumber\\
A_{0}=\frac{p_{x_{0}}}{\omega_{1}}\sqrt{1+\frac{\omega_{1}^{2}\omega_{2}^{2}\theta^{2}
x_{0}^{2}}
{4(1+\omega_{2}^{2}u_{0}^{2})}}
\end{eqnarray}

In this way we have shown that using a gauge transformation induced in the
complex space. We can solve different systems starting from the solution
of the harmonic oscillator.  Of course, the kind of systems
that we can solve depend from the starting complex model and the gauge
conditions imposed.

In next two sections, we want to implement these ideas to the quantum
level, first using the path integral in the Batalin-Fradkin-Vilkovisky
formalism BFV \cite{BFV1} and after that to the canonical level.

\section{BFV Formalism In The Harmonic Oscillator In One
Dimension}\label{brst}
In this Section, we show at the quantum level, a relationship that exist
between different gauge conditions and the way in that changes
the corresponding quantum theory. To manage this problem  we introduce the
BRST formalism \cite{BFV}, in the formulation of BFV using as an example
the complex harmonic oscillator in one
dimension.

First of all,  consider a  formalism with one complex dimension
and start from the Lagrangian
\begin{equation}\label{CHO.1}
 L=\frac{1}{2}\dot{z}^{2}-\frac{\omega^{2}}{2}z^{2}
\end{equation}
where $z$ is a complex variable that is separated into real and imaginary
parts
\begin{equation}
 z=x+iy.
\end{equation}
The Lagrangian in terms of these variables is
\begin{eqnarray}\label{Lagrangianacom.11}
 L_{(x,y)}=\frac{1}{2}\dot{x}^{2}-\frac{1}{2}\dot{y}^{2}+i\dot{x}\dot{y}-\frac{\omega^{2}}{2}x^{2}+\frac{\omega^{2}}{2}y^{2}
-i\omega^{2}xy
\end{eqnarray}
and we obtain the momenta
\begin{eqnarray}
 p_{x}=\dot{x}+i\dot{y},\\
p_{y}=-\dot{y}+i\dot{x}.
\end{eqnarray}
We observe that these momenta are not independent, then generate the
primary constraint
\begin{equation}\label{cpc}
 \Phi_{0}=p_{x}+ip_{y}\approx 0.
\end{equation}
The next step is to obtain the canonical Hamiltonian
\begin{equation}
 H_{0}=\dot{x}p_{x}+\dot{y}p_{y}-L_{(x,y)},
\end{equation}
and the total Hamiltonian will be
\begin{equation}\label{oaham.1}
 H_{T}=\frac{1}{2}p_{x}^{2}+\frac{\omega^{2}}{2}x^{2}-\frac{\omega^{2}}{2}y^{2}+i\omega^{2}xy+\mu^{0}
\Phi_{0}.
\end{equation}
From the temporal evolution of (\ref{cpc}) with the total Hamiltonian
(\ref{oaham.1}) we observe  that the primary constraint is of first class
and is the generator of a gauge symmetry.

In order to reduce the above theory, we must eliminate degrees of freedom
by fixing a gauge condition that in general is a complex
function, but its election is not unique.

Specifically, we choose two ways to fix the gauge
$\gamma_{0,1}$ with
\begin{eqnarray}\label{gauge1d}
\gamma_{0}=y\approx0,\nonumber\\
\gamma_{1}=y-ix+iU^{\frac{1}{2}}(x),
\end{eqnarray}
where the first election implies a real formulation of a harmonic
oscillator. In contrast to the second gauge condition, that permits non
trivial results
\begin{eqnarray}
\{\Phi,\gamma_{1}\}=-\frac{i}{2U^{\frac{1}{2}}}\partial_{x}U(x).\nonumber
\end{eqnarray}
Now we can use the BRST formalism. Here the phase space is extended
including fermionic degrees of freedom. With these degrees of freedom, we
establish the BRST symmetry that is a
global remnant of a local gauge symmetry after the gauge fixing.
 The importance of a theory with these characteristics
is based in that the propagator is a sum over all probabilities
of process that involves only real particles on initial and final states
and in consequence is unitary.

As the first step, we consider that  the Lagrange multiplier $\mu^{0}$ ,
in (\ref{oaham.1}), is a variable
of the configuration space with its respective momenta
$\pi_{0}$. For these   variables, we associate the Poisson brackets
\begin{eqnarray}
\{\mu^{0},\pi_{0}\}=-\{\pi_{0},\mu^{0}\}=1,
\end{eqnarray}
which is part of an extended symplectic structure.

Additionally, it is necessary to include in the phase space
a ghost $\mathcal{C}$ and an anti-ghost $\bar {\mathcal{C}}$ with
symplectic structure given by
\begin{eqnarray}
 \{\bar{\mathcal{P}},\mathcal{C}\}= \{\mathcal{P},\bar{\mathcal{C}}\}=-1,
\end{eqnarray}
and zero for another brackets. From these ghost and anti-ghost, we
obtain the BRST charge, since
can be seen that exists a new symmetry in this extended theory
\begin{eqnarray}
\Omega=\mathcal{C}\Phi_{0}-i\mathcal{P}\pi_{0},
\end{eqnarray}
and the BRST charge $\Omega$ is the generator of the BRST symmetry.
Furthermore, the gauge conditions are included through the fermionic
conditions
\begin{eqnarray}\label{f.conditions}
\Psi_{0}=i\bar{\mathcal{C}}\gamma_{0}+\bar{\mathcal{P}}\mu^{0},\qquad
\Psi_{1}=i\bar{\mathcal{C}}\gamma_{1}+\bar{\mathcal{P}}\mu^{0},
\end{eqnarray}
resulting from the ghosts.
Using these conditions and the BRST charge, we obtain a brackets that
are used later to obtain the effective Hamiltonian
\begin{eqnarray}
 \{\Psi_{0},\Omega\}=-\mu^{0}\Phi_{0}+\pi_{0}\gamma_{0}-i\bar{\mathcal{P}}\mathcal{P}
-\bar{\mathcal{C}}\mathcal{C},\\
 \{\Psi_{1},\Omega\}=-\mu^{0}\Phi_{0}+\pi_{0}\gamma_{1}-i\bar{\mathcal{P}}\mathcal{P}
-\bar{\mathcal{C}}\mathcal{C}(\frac{\partial_{x}U}{2U^{\frac{1}{2}}}).\nonumber
\end{eqnarray}
From this structure is obtained a way to achieve the quantization by means
of the path integral. Before it is necessary to show that exist a gauge
freedom in the path integral and it is independent from the fermionic
conditions (\ref{f.conditions}), {\it i.e.} the Fradkin-Vilkovisky theorem \cite{BFV}.

The path integral using the BRST formalism and applied to the complex
theory with a gauge condition $\Psi_{0}$ is
\begin{eqnarray}\label{intcambrst}
 Z_{\Psi}=\int \!\mathcal{D}\mu \exp(iS_{eff}),\\
S_{eff}=\int_{\tau_{1}}^{\tau_{2}}\! d\tau
(\dot{x}p_{x}+\dot{y}p_{y}-\mu^{0}\dot{\pi}_{0}
\nonumber\\
+\bar{\mathcal{C}}\dot{\mathcal{P}}+\bar{\mathcal{P}}\dot{\mathcal{C}}
-H_{eff}),\\
H_{eff}=H_{BRST}-\{\Psi_{0},\Omega\}
,\nonumber\\
H_{BRST}=H_{0},\\
H_{eff}=\frac{1}{2}p_{x}^{2}+\frac{\omega^{2}}{2}x^{2}-\frac{\omega^{2}}{2}y^{2}+i\omega^{2}xy
+\bar{\mathcal{C}}\mathcal{C}\nonumber\\
+\mu^{0}\Phi_{0}-\pi_{0}\gamma_{0}+i\bar{\mathcal{P}}\mathcal{P},\\
d\mu=dxdp_{x}dydp_{y}d\mu^{0}d\pi_{0}d\mathcal{P}d\mathcal{C}d\bar{\mathcal{P}}d\bar{\mathcal{C}},
\end{eqnarray}
where it is possible to establish variations and to compare
between Fermionic conditions
$\Psi_{0}$ and $\Psi_{1}$. As the starting point we establish
a infinitesimal difference that is
\begin{eqnarray}
\chi_{0}=i\int_{t_{1}}^{t_{2}}\!d\tau(\Psi_{1}-\Psi_{0})=i\int_{t_{1}}^{t_{2}}\!d\tau\bar{\mathcal{C}}
(x-U^{\frac{1}{2}}(x)).
\end{eqnarray}
With this above difference is possible to establish a transformation that
change variables resulting
\begin{eqnarray}
 x'=x+\{x,(\mathcal{C}\Phi_{0}-i\mathcal{P}\pi_{0})\}\chi_{0}=x+\mathcal{C}\chi_{0},\nonumber\\
y'=y+\{y,(\mathcal{C}\Phi_{0}-i\mathcal{P}\pi_{0})\}\chi_{0}=y+i\mathcal{C}\chi_{0},\nonumber\\
\mu'^{0}=\mu^{0}+\{\mu^{0},(\mathcal{C}\Phi_{0}-i\mathcal{P}\pi_{0})\}\chi_{0}
=\mu^{0}-i\mathcal{P}\chi_{0},\nonumber\\
p_{x'}=p_{x}+\{p_{x},(\mathcal{C}\Phi_{0}-i\mathcal{P}\pi_{0})\}\chi_{0}=p_{x},\nonumber\\
p_{y'}=p_{y}+\{p_{y},(\mathcal{C}\Phi_{0}-i\mathcal{P}\pi_{0})\}\chi_{0}=p_{y},\nonumber\\
\pi'_{0}=\pi_{0},\nonumber\\
-i\mathcal{P}'=-i\mathcal{P}+\{-i\mathcal{P},(\mathcal{C}\Phi_{0}-i\mathcal{P}\pi_{0})\}\chi_{0}=
-i\mathcal{P},\nonumber\\
\mathcal{C}'=\mathcal{C}+\{\mathcal{C},(\mathcal{C}\Phi_{0}-i\mathcal{P}\pi_{0})\}\chi_{0}=\mathcal{C}\nonumber\\
\bar{\mathcal{C}}'=\bar{\mathcal{C}}+\{\bar{\mathcal{C}},(\mathcal{C}\Phi_{0}-i\mathcal{P}\pi_{0})\}\chi_{0}
=\bar{\mathcal{C}}+i\chi_{0}\pi_{0},
\nonumber\\
\bar{\mathcal{P}}'=\bar{\mathcal{P}}+\chi_{0}\{\bar{\mathcal{P}},(\mathcal{C}\Phi_{0}-i\mathcal{P}\pi_{0})\}
=\bar{\mathcal{P}}-\chi_{0}\Phi_{0},\label{brsttransfomation}
\end{eqnarray}
and that is applied to the BRST Hamiltonian
\begin{eqnarray}
 H'_{BRST}=\frac{1}{2}p_{x}^{2}+\frac{\omega^{2}}{2}(x+iy)^{2}
 =H_{BRST},
\end{eqnarray}
that is invariant under the gauge transformations
(\ref{brsttransfomation}).

On another side, the kinetic term is invariant under these BRST
transformations
\begin{eqnarray}
 \dot{x}'p'_{x}+\dot{y}'p'_{y}-\mu'^{0}\dot{\pi}'_{0}
+\bar{\mathcal{C}}'\dot{\mathcal{P}}'+\bar{\mathcal{P}}'\dot{\mathcal{C}}'=\nonumber\\
 \dot{x}p_{x}+\dot{y}p_{y}-\mu^{0}\dot{\pi}_{0}
+\bar{\mathcal{C}}\dot{\mathcal{P}}+\bar{\mathcal{P}}\dot{\mathcal{C}}.
\end{eqnarray}
So, in order to show that the path integral is invariant only is necessary
to analyze a part of the effective Hamiltonian and the measure of the path
integral.

Because, the kinetic part is invariant now is necessary to find
as the effective Hamiltonian changes. The variation that results
from these transformations in the effective Hamiltonian is
\begin{equation}
\delta H_{eff}=\{(\Psi_{1}-\Psi_{0}),\Omega\}=\mathcal{C}
\bar{\mathcal{C}}
(1-\frac{\partial U^{\frac{1}{2}}}
{\partial x})+
i\pi_{0}(x-U^{\frac{1}{2}})])
\end{equation}
where were used the gauge conditions  (\ref{gauge1d}).

The change in the effective action is
\begin{eqnarray}
e^{i\delta
S_{eff}}=\exp{\left(-i\int\!d\tau\{\Omega,(\Psi_{1}-\Psi_{0})\}\right)}\nonumber\\
=\exp{\left(-i\int\!d\tau[\mathcal{C}\bar{\mathcal{C}}(1-\frac{\partial
U^{\frac{1}{2}}}
{\partial x})+i\pi_{0}(x-U^{\frac{1}{2}})]\right)},
\end{eqnarray}
where it is shown  an extra term that is originated from the proposed change.

On another side, it is important to consider the measure of
the path integral. Here we are considered an infinitesimal change in
the variables produced by these gauge transformations
\begin{eqnarray}
 \frac{\delta x'}{\delta
x}=\delta(t-t')+i\mathcal{C}\bar{\mathcal{C}}(1-\partial_{x}U^{\frac{1}{2}}
),\\
\frac{\delta \bar{\mathcal{C}}'}{\delta
\bar{\mathcal{C}}}=\delta(t-t')-\pi_{0}(x-U^{\frac{1}{2}}),\nonumber
\end{eqnarray}
and other variations are a Dirac delta that depends on time.
With these variations the change for the measure of the  path integral is
\begin{eqnarray}
\mathcal{D}\mu'=\mathcal{D}\mu \exp{\left(i\int\!
d\tau[\mathcal{C}\bar{\mathcal{C}}(1-\frac{\partial
U^{\frac{1}{2}}}
{\partial x})+i\pi_{0}(x-U^{\frac{1}{2}})]\right)}.
\end{eqnarray}

And now it is possible to consider all above elements and to apply
in the path integral, in a way infinitesimal, in order
to show that the path integral is invariant under the change generated by
this symmetry
\begin{eqnarray}
Z_{\Psi_{0}}=\int \!\mathcal{D}\mu \exp(iS_{eff})
=\int \!\mathcal{D}\mu' \exp(iS'_{eff})=Z_{\Psi_{1}}.
\end{eqnarray}

It is important to mention that the BRST charge generates an infinitesimal
change in the phase space that if we apply to the elements of the path
integral (\ref{intcambrst}). We observe a contribution that helps to
pass from a Fermionic condition to another. So, we have shown the
Fradkin-Vilkovisky theorem for this
complex case and we have shown  that both systems with gauge conditions
$\gamma_{0}$ and $\gamma_{1}$ are connected.
In the next Section, we calculate the path integral for
the complex harmonic oscillator in one dimension.

\section{Path integral and Gauge Conditions}\label{path}
It is a central point to establish a quantum description of this
complex model and it is important to include the gauge freedom. A way
to achieve this description is through the Senjanovic method that
allows to  quantize systems with second class constraints
\cite{Senjanovic}. In a particular form, we return to harmonic oscilllator
in two dimensions (\ref{hamred})  with gauge coditions
$\gamma_{0*}$, $\gamma_{1*}$  (\ref{nornc}) and the first class
constraints (\ref{phi0}) that are introduced using the next notation
\begin{eqnarray}
 \xi^{a}=(x, y,u,v),\qquad \rho_{a}=(p_{x}, p_{y}, p_{u}, p_{v}),\\
\tau_{b}=(\Phi_{0},\gamma_{0*}, \Phi_{1} ,\gamma_{1*})
\end{eqnarray}
where $\xi^{a}$ includes the real and imaginary parts of
$z_{1}$, $z_{2}$ and $\rho_{a}$ that is a complex quantity
for momenta.

From these elements now  it is possible to present the path integral that
includes first class constraints and gauge conditions. The integration
measure is
\begin{eqnarray}
 \mathcal{D}\Xi=\mathcal{D}\xi^{a}\mathcal{D}\rho_{a}\det\mid\left\lbrace\tau_{b},\tau_{c}\right\rbrace\mid
\delta(\Phi_{0})\delta(\gamma_{0*})\delta(\Phi_{1})
\delta(\gamma_{1*}),
\end{eqnarray}
and the total patgh integral is
\begin{eqnarray}
Z=\int\mathcal{D}\Xi \exp [i\int\!dt
(\dot \xi^a \rho_a-H_{2C})].
\end{eqnarray}
From the path integral and these constraints, it is possible
to eliminate degrees of freedom $(y,p_{y}, v,p_{v})$, that are
imaginary parts through the delta functions $\delta(\tau_{C})$, resulting
\begin{eqnarray}
Z_{R}=\int\mathcal{D} x\mathcal{D} p_{x}\mathcal{D} u\mathcal{D} p_{u}
\prod_{j}|p_{x^{j}}p_{u^{j}}|^{2}\nonumber\\
\exp [i\int\!dt
(-p^{2}_{x}\dot{x}-xp_{x}\dot{p}_{x}
-p^{2}_{u}\dot{u}-up_{u}\dot{p}_{u}-H_{1Cons})]
\end{eqnarray}
that is the real reduced path integral.

It is interesting to mention as we have achieved to reduce a complex
formalism in a real theory by means of a particular way
of projecting. This projection is not unique and exists an infinite number of possibilities according to the gauge conditions selected.

\section{ Canonical quantization}\label{canonical}

To analyze the role played by this gauge symmetry in the canonical
quantization we parametrize the action for the model introduced in
(\ref{CHO.1}). In this case the action
is transformed to
\begin{equation}
S=\int_{\tau_0}^{\tau_1} d\tau \left( \frac{\dot z^2}{2\dot
t}-\frac{\omega^2}{2}\dot t z^2\right)
\end{equation}
in terms of real and imaginary parts we get
\begin{equation}
S=\int_{\tau_0}^{\tau_1} d\tau \left( \frac{(\dot x +i \dot y)^2}{2\dot
t}-\frac{\omega^2}{2}\dot t (x+iy)^2\right)
\end{equation}
Notice that we are assuming that the time variable $t$ is a real variable.
For  the momenta we obtain
\begin{equation}
p_x= \frac{(\dot x +i \dot y)}{\dot t}, \qquad p_y= \frac{(i \dot x - \dot
y)}{\dot t}, \qquad p_t =-\frac{(\dot x +i \dot y)^2}{2\dot
t^2}-\frac{\omega^2}{2}(x+iy)^2  \end{equation}

From these expressions we see that there are now  two constraints
\begin{eqnarray}
\Phi&=& p_x+ip_y\approx 0, \\
\Psi&=& p_t +\frac{p_x^2}{2} + \frac{\omega^2}{2} (x+iy)^2
\approx 0.
\end{eqnarray}
By computing the canonical Hamiltonian we see that vanishes then the total
Hamiltonian is given by
\begin{equation}
H_T=\mu^1 \Phi + \mu_2 \Psi,
\end{equation}
From the evolution of the constraints we find that there are not more
constraints and that both constraints are first class.
To quantize this system we will fix partially the gauge, using a gauge
condition for the constraint $\Phi$.  For the other first class constraint
$\Psi$,
we will use Dirac's quantization condition
\begin{equation}
\hat \Psi |\psi \rangle =0,
\end{equation}
For a given basis $|z,t\rangle$, we get
\begin{equation}
\left(\hat p_t +\frac{\hat p_x^2}{2} + \frac{\omega^2}{2} (\hat x+i\hat
y)^2\right)\psi(x+iy,t)=0.
\end{equation}
For example, fixing the gauge
\begin{equation}\label{G101}
\gamma_1 = y\approx 0,
\end{equation}
 we obtain the usual harmonic oscillator, with Dirac's brackets
\begin{equation}\label{DB106}
\{x,p_x\}^*=1,
\end{equation}
and normalized eigenfunctions
\begin{equation}\label{EF103}
\psi(x,t)=\frac{1}{\sqrt{2^n n!}}\left(\frac{\omega}{\pi
\hbar}\right)^{1/4}H_n\left(\sqrt{\frac{\omega}{\hbar}}x\right)\exp\left(-\frac{\omega
x^2}{2\hbar}\right)\exp\left(\frac{-i}{\hbar}E_n t\right)
\end{equation}
But now, fixing a different gauge, for example
\begin{equation}
\gamma_2=y-i\left(x-U^{1/2}(x)\right)\approx 0,
\end{equation}
In this case the constraint $\Psi$ is reduced to
\begin{equation}
\Psi=\left(\hat p_t +\frac{\hat p_x^2}{2} + \frac{\omega^2}{2} U(\hat
x)\right)\approx 0,\end{equation}
and the Dirac bracket for the reduced space is given by
\begin{equation}
\{x,p_x\}^*=\frac{2 U^{1/2}}{\partial_x U},\end{equation}
we can check that the above bracket is equal to (\ref{DB106}) in the
case of $U(x)=x^2$.
By promoting this bracket to commutators in the quantum theory, the
momentum operator must be realized as
\begin{equation}
p_x=-i\hbar \frac{2 U^{1/2}}{\partial_x U}\partial_x. \end{equation}
In consequence our Schr\"odinger equation is given explicitly by
 \begin{equation}\label{S108}
\left(-i\hbar \frac{\partial}{\partial t}-\frac{\hbar^2}{2} \left(\frac{2
U^{1/2}}{\partial_x U}\partial_x \right)^2+ \frac{\omega^2}{2} U(\hat
x)\right)\psi(U^{1/2},t)=0.\end{equation}
One can easily check that the eigenfunctions (\ref{EF103}) with the
substitution $x\to U^{1/2}$ are solutions of the equation (\ref{S108}) and
in this sense the problem with the gauge (\ref{G101}) (the harmonic
oscillator) is gauge related to the problem (\ref{S108}). In this form, we have shown 
that for different gauge choices the quantum problems are gauge related, also in the quantum canonical sense.

\section{Discussion and Conclusions}\label{conclu}
In this work, we introduce a new symmetry for complex holomorphic systems. This invariance is a local symmetry 
that it is trivial when is written in terms of complex variables. However, by written it in terms of real and imaginary parts allows us to 
connect in the complex space, real systems that apparently are not connected.  

The general idea of the proposed formalism is to write any real theory in terms of complex variables, then to show that there is a gauge symmetry of the system.
The next step is to analyze this symmetry in the canonical formalism and show that associated with the gauge symmetry we have first class constraints.
This implies that we need to fix the gauge, in consequence we can select any "good" canonical gauge to project  from the complex space to the real space.  
For each of the gauge choices,  we have a different theory in real space. However, each of these theories is related by a gauge transformation in the complex space.
The validity interval of each gauge condition is determined by the Dirac theory of constraints \cite{Dir}. Finally, we show using canonical quantization, that the solutions of the respective Schr\"odinger equations, of two systems connected by a gauge transformation are also gauge related.

In this way, in the present work we have shown that following the idea developed in \cite{margalli:gnus}, we can consider different types of gauge conditions that lead us to different real systems. Furthermore,  that using the gauge transformations we can find the solutions of the systems that are  gauge connected, both at the classical level as at the quantum level. In some sense our procedure, is a generalization of the canonical transformations. Since for a given gauge condition we can get a system with a different symplectic structure. Also, in some sense, is related to the two-time physics of Bars \cite{Bars}, where from a space with two times we can get by using a gauge choice different one time systems. The essential difference is that in our case  we have more freedom, since we can choose any real system,  write it in complex coordinates and then project it in different ways to real space. Our procedure can be generalized to field theory and also to high order  time derivative theories \cite{margalli2}.

\section{Acknowledgments }
The authors acknowledge partial support from CONACYT project 237503 and
DGAPA-UNAM grant IN 103716.


\section*{References}

\begin{thebibliography}{00}
\bibitem{Dir}  Dirac P A M 2001 Lectures on Quantum Mechanics
(vol 151) ed Dover (New York) chapter 1 pp 1-25

\bibitem{Hen} Henneaux M and  Teitelboim C 1992 Quantization of Gauge
Systems ed Princeton University Press (Princeton)

\bibitem{Gribov} Gribov V N 1978 Quantization of non-Abelian gauge
theories {\it Nucl. Phys. B} {\bf 139} 1

\bibitem{Ashtekar.b} Ashtekar A 1991 Lectures on Non-perturbative
Canonical Gravity ed World Scientific (Singapore)

\bibitem{Bender.1}  Bender C M  and  Boettcher S 1998 {\it Phys. Rev.
Lett.} {\bf 80} 5243

\bibitem{Moise}  Moiseyev N 2011  Non-Hermitian Quantum Mechanics ed
Cambridge U P (Cambridge)


\bibitem{Galileo} Gilmore R 1974 Lie Groups, Lie Algebras and some of
their applications ed J. Wiley (New York)

\bibitem{margalli:gnus}
Margalli C A and   Vergara J D 2015  Hidden Gauge Symmetry in
Holomorphic Models {\it Phys.Lett. A} {\bf 379}  2434

\bibitem{Moshinsky} Moshinsky M and  Quesne C 1971 Linear Canonical
Transformations and Their Unitary Representations {\it J. Math. Phys.}
{\bf 12} 1772

\bibitem{Morales} Morales-Tecotl H A,  Urrutia L F and  Vergara J D
1996 {\it Class.\ Quant. Grav.} {\bf 13} 2933 {\it Preprint} gr-qc/9607044

 \bibitem{complex}R. Remmert 1991 Theory of Complex Functions ed
Springer-Verlag (New York)

\bibitem{BFV} Heneaux M 1985 Hamiltonian form of the path integral for
theories with gauge freedom {\it Phys. Rep.} {\bf 126}1

\bibitem{BFV1} Fradkin E S and  Vilkovisky G A 1977  {\it CERN Report
TH-2332}

\bibitem{Senjanovic} Senjanovic P 1976 {\it Ann. of Phys.} {\bf 100} 227

\bibitem{Bars} Bars I 2001 {\it Class. Quantum Grav.} {\bf18} 3113

\bibitem{margalli2} Margalli C A, Vergara J D and Romero J M 2017 work in progress.




\end{thebibliography}
\end{document}